\pdfoutput=1

\documentclass[12pt]{article}
\usepackage{graphicx}
\usepackage{hyperref}
\usepackage{amsmath,booktabs,array}
\usepackage{mathtools}
\usepackage{lineno}
\def\institute{Department of Physics\\University of Liverpool, UK}
\def\support{\footnote{On behalf of the LHCb collaboration}}

\def\Title#1{\begin{center} {\Large #1 } \end{center}}
\def\Author#1{\begin{center}{ \sc #1} \end{center}}
\def\Address#1{\begin{center}{ \it #1} \end{center}}

\newenvironment{Abstract}{\begin{quotation}  }{\end{quotation}}
\newenvironment{Presented}{\begin{quotation} \begin{center} 
             PRESENTED AT\end{center}\bigskip 
      \begin{center}\begin{large}}{\end{large}\end{center} \end{quotation}}

\def\beq{\begin{equation}}
\def\eeq#1{\label{#1}\end{equation}}
\def\eeqn{\end{equation}}

\def\beqa{\begin{eqnarray}}
\def\eeqa#1{\label{#1}\end{eqnarray}}
\def\eeqan{\end{eqnarray}}

\let\bar=\overbar

\def\Dslash{\not{\hbox{\kern-4pt $D$}}}
\def\dslash{\not{\hbox{\kern-2pt $\del$}}}

\def\msb{{\bar{\ssstyle M \kern -1pt S}}}

\begin{document}
\begin{titlepage}
%\pubblock

\vfill
\Title{Tops in the forward region}
\vfill
\Author{ James V Mead\support}
\Address{\institute}
\vfill
\begin{Abstract}
Detailed results on top quark production at LHCb are reported from the first observation of top production in a forward region detector to the first measurement of top pair production at LHCb with 13 TeV data.
\end{Abstract}
\vfill
\begin{Presented}
\href{https://indico.cern.ch/event/690229/contributions/2941634/}{$11^\mathrm{th}$ International Workshop on Top Quark Physics\\
Bad Neuenahr, Germany, September 16--21, 2018}
\end{Presented}
\vfill
\end{titlepage}
\def\thefootnote{\fnsymbol{footnote}}
\setcounter{footnote}{0}

%\linenumbers

\section{Introduction}

The LHCb detector's unique coverage extends access to new kinematic regions of high energy particle interactions. In spite of its comparatively small acceptance, LHCb has advanced particle identification and forward tracking allowing precise measurements of top production in the forward region. The vertex locator (VELO), surrounding the interaction point, provides vertex reconstruction capabilities of particular benefit to heavy flavour (HF) tagging in jets. LHCb also offers a low pile-up environment, a sub percent fake jet rate and \textit{b}-jet mistag rate of $\sim\,0.3\%$ \cite{bcjets}.

In probing extremes of phase space in the $x$-$Q^{2}$ plane, LHCb offers significant improvements to uncertainties surrounding the contents of the proton \cite{WZjets}. The gluon parton distribution function (PDF) is maximally correlated to the top pair cross-section at high-$x$, where it is poorly constrained. The top quark mass makes their production a natural probe of this region \cite{gPDFconstrains}. Forward top production in particular may, therefore, constrain the \textit{g}-PDF by $\mathcal{O}(20\%)$ \cite{toplargex}.

The Standard Model (SM) predicts that interference effects at next-to-leading-order (NLO) produce an asymmetry in top pairs from quark-initiated production. The dilution of this asymmetry by the dominant gluon fusion process reduces in LHCb's forward acceptance resulting in a larger asymmetry with respect to the central region \cite{topasymm}. With sufficient statistics, prospects improving further for future runs, LHCb is in place to contribute to the LHC's ongoing top program \cite{upgrade}. \\\\\\\\

%%%%%%%%%%%%%%%%%%%%%%%%%%%%%%%%%%%%%%%%%%%%%%%%%%%%%%%%%%%%%%%%%%%%%%%%%
%%
%%   use this format to include a LaTeX table  into your paper
%%
\begingroup
\setlength{\tabcolsep}{15pt} % Default value: 6pt
\begin{table}[h]
\begin{center}
\begin{tabular}{l|r<{}@{\;$\pm$\;}>{}lr<{}@{\;$\pm$\;}>{}lr<{}@{\;$\pm$\;}>{}l}

$d\sigma($fb$)$ 
&  \multicolumn{2}{c}{$7\,$TeV}   
&  \multicolumn{2}{c}{$8\,$TeV} 
&  \multicolumn{2}{c}{$14\,$TeV} \\ \hline

$ lb $      &   285 \hspace{.2cm} & \hspace{.35cm} 52     &     504 \hspace{.2cm} & \hspace{.35cm} 94     &    4366 \hspace{.2cm} & \hspace{.35cm} 663   \\
$ lbj $     &   97 \hspace{.2cm} & \hspace{.35cm} 21      &     198 \hspace{.2cm} & \hspace{.35cm} 35     &    2335 \hspace{.2cm} & \hspace{.35cm} 323   \\
$ lbb $     &   32 \hspace{.2cm} & \hspace{.35cm} 6       &     65 \hspace{.2cm} & \hspace{.35cm} 12      &    870 \hspace{.2cm} & \hspace{.35cm} 116    \\
$ lbbj $    &   10 \hspace{.2cm} & \hspace{.35cm} 2       &     26 \hspace{.2cm} & \hspace{.35cm} 4       &    487 \hspace{.2cm} & \hspace{.35cm} 76     \\
$ l^{+}l^{-} $ & 44 \hspace{.2cm} & \hspace{.35cm} 9       &     79 \hspace{.2cm} & \hspace{.35cm} 15      &    635 \hspace{.2cm} & \hspace{.35cm} 109    \\
$ l^{+}l^{-}b $ & 19 \hspace{.2cm} & \hspace{.35cm} 4       &     39 \hspace{.2cm} & \hspace{.35cm} 8       &    417 \hspace{.2cm} & \hspace{.35cm} 79     \\ \hline

\end{tabular}
\caption{$t\bar{t}$ cross-section channels within LHCb's acceptance with quoted uncertainty accounting for variation of scale, PDF and shower modelling uncertainty \cite{topasymm}.}
\label{tab:blood}
\end{center}
\end{table}
\endgroup
%%%%%%%%%%%%%%%%%%%%%%%%%%%%%%%%%%%%%%%%%%%%%%%%%%%%%%%%%%%%%%%%%%%%%%%%%%%

\section{Results}

The reconstruction of top decays typically involves the production of $Wb$ with a signature of 3 jets or a lepton$\,+\,$jet per top quark. The following results on forward region top production, measured with the LHCb detector, are from the: semi-leptonic channel using $7\,\&\,8\,$TeV data; di-jet channel using $8\,$TeV data; di-lepton channel using $13\,$TeV data.

\subsection{First observation, $\mu + b$ final state}

Electroweak (EW) boson and associated jet measurements are performed using data corresponding to 1.0 and 2.0$\,$fb$^{-1}$ of integrated luminosity collected in Run I at 7 and 8$\,$TeV. Minimum p$_{T}$ requirements on the $\mu$ and $b$-jet help limit multi-jet QCD and dominant $Wb$ backgrounds respectively. The acceptance for jets is reduced at the fringes of the detector to negate edge effects and ensure flat reconstruction efficiency. A minimum requirement on the vector sum of muon and jet p$_{T}$ approximates the missing $E_{T}$ of imbalanced EW events. A data driven template for the QCD contribution is taken from the p$_{T}$ balanced control region. A profile likelihood fit may be performed to establish the presence of top, then, having subtracted simulated $\sigma(Wb)/\sigma(Wj)$ normalised to data, the cross-section may be measured.

%%%%%%%%%%%%%%%%%%%%%%%%%%%%%%%%%%%%%%%%%%%%%%%%%%%%%%%%%%%%%%%%%%%%%%%%%
%%
%%   use this format to include an .eps figure into your paper
%%
\begin{figure}[htb]
\centering
\includegraphics[height=2.1in]{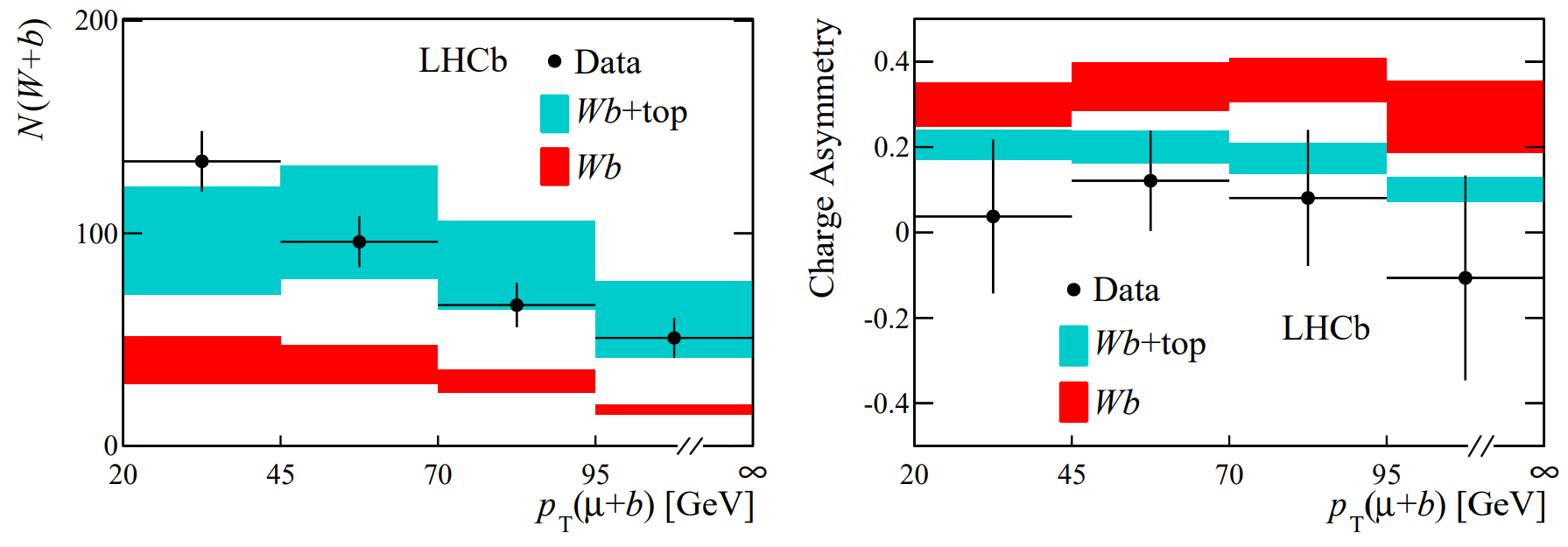}
\caption{Results for the $W+b$ yield (left) and charge asymmetry (right) versus p$_{T}(\mu + b)$
compared to SM predictions obtained at NLO using MCFM \cite{firstobs} .}
\end{figure}
%%%%%%%%%%%%%%%%%%%%%%%%%%%%%%%%%%%%%%%%%%%%%%%%%%%%%%%%%%%%%%%%%%%%%%%%%%%
The resulting inclusive top production cross-sections, observed to 5.4$\sigma$, in the fiducial region defined by: p$_{T}(\mu)>25\,$GeV; $2.0<\eta(\mu)<4.5\,$; $50<$p$_{T}(b)<100\,$GeV; $2.2<\eta(b)<4.2$; $\Delta$R$(\mu,b)>0.5$; and p$_{T}(\mu+b)>20\,$GeV are:\\

\begingroup
\setlength{\tabcolsep}{.03\linewidth}
\begin{tabular}{rr}
    $\sigma$(top)[7\,TeV] \hspace{.2cm} = \hspace{.2cm} 239 $\pm$ 53 (stat) $\pm$ 33 (syst) $\pm$ 24 (theory) fb , & \\
    $\sigma$(top)[8\,TeV] \hspace{.2cm} = \hspace{.2cm} 289 $\pm$ 43 (stat) $\pm$ 40 (syst) $\pm$ 29 (theory) fb . & \cite{firstobs} 
\end{tabular}
\endgroup

\vspace{.3cm}
\noindent The uncertainty is currently dominated by the $\it{b}$-tagging efficiency. These results, including differential yields and charge asymmetries, are in agreement with SM predictions at NLO.

\subsection{Pair production, $l + bb$ final state}

A simultaneous four-dimensional fit to $8\,$TeV data, corresponding to an integrated luminosity of 2$\,$fb$^{-1}$, provides access to the $t\bar{t}$, $W+bb$ and $W+cc$ cross-sections. Using data split by lepton flavour and charge, the fit is performed to: the HF boosted decision tree (BDT) output for each jet; the di-jet invariant mass; the gradient boosted MVA output, \textsc{uGB}, with minimised correlation with invariant mass and trained to discriminate between the three signal channels. The \textsc{uGB} itself is trained on p$_{T}$, $\eta$ and $\Delta$R of the decay products as well as the jet masses and $cos(\theta_{l})$, where $\theta_{l}$ is the lepton scaterring angle in the di-jet rest frame.

%%%%%%%%%%%%%%%%%%%%%%%%%%%%%%%%%%%%%%%%%%%%%%%%%%%%%%%%%%%%%%%%%%%%%%%%%
%%
%%   use this format to include an .eps figure into your paper
%%
\begin{figure}[htb]
\centering
\includegraphics[height=2.5in]{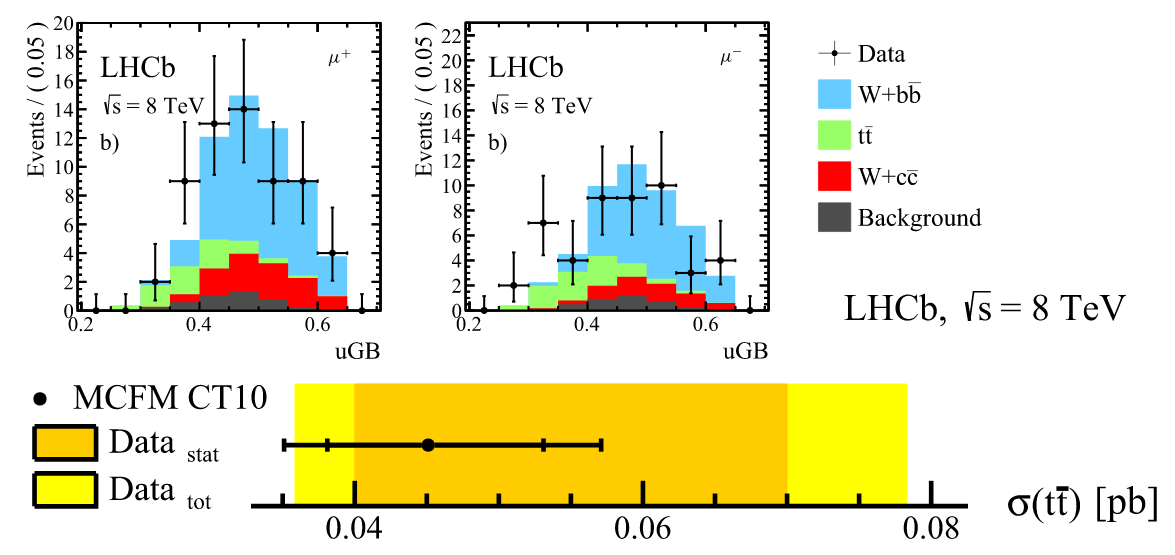}
\caption{Result of simultaneous 4D fit in terms of \textsc{uGB} response for $\mu^{+}$ (left) and $\mu^{-}$ (right); observed and expected cross-sections compared, where inner uncertainties on theory represent the scale errors (bottom) \cite{dijet}.}
\end{figure}
%%%%%%%%%%%%%%%%%%%%%%%%%%%%%%%%%%%%%%%%%%%%%%%%%%%%%%%%%%%%%%%%%%%%%%%%%%%

The resulting inclusive $t\bar{t}$ production cross-section in the fiducial region defined by p$_{T}(\mu,e)>(20,15)\,$GeV; $2.0<\eta(\mu,e)<(4.5,4.25)\,$; $12.5<$p$_{T}(j)<100\,$GeV; $2.2<\eta(j)<4.2$; $\Delta$R$(l,j)>0.5$; $\Delta$R$(j_{1},j_{2})>0.5$; p$_{T}(l+j_{1}+j_{2})>15\,$GeV is:\\

\begingroup
\setlength{\tabcolsep}{.15\linewidth}
\begin{tabular}{r}
    $\sigma$($t\bar{t}$)[8\,TeV] \hspace{.2cm} = \hspace{.2cm} 0.05 $_{-0.01}^{+0.02}$ (stat) $_{-0.01}^{+0.02}$ (syst) pb \cite{dijet}.
\end{tabular}
\endgroup

\vspace{.3cm}
\noindent These results are in agreement with SM predictions and imply $t\bar{t}$ observation to 4.9$\sigma$ significance. With precision limited, both, systematically and by low statistics, this measurement looks to benefit from future data sets with which background modelling and systematic uncertainties may be improved.

\subsection{Run II, $\mu e + b$ final state}

The greater centre of mass energy of Run$\,$II provides an increased yield of $t\bar{t}$ with up to a factor of 10 gained in LHCb's acceptance. With just 2$\,$fb$^{-1}$ of data, the purest channel, previously inaccessible, becomes viable for analysis. Angular separation requirements, lepton isolation and impact parameter (IP) cuts are all applied.

The inclusion of a second lepton suppresses both $Wb$ and multi-jet QCD while the different flavours suppress the $Z+$jet background. From simulation: single top becomes an irreducible $Wt$ background scaled to theory prediction; $Zj$ is scaled using the $Z$ peak in data; estimates of leptons from mis-ID are taken from data. A background subtraction is performed and simulated $t\bar{t}$ is normalised to the remainder of the 44 events in data.  \\

%%%%%%%%%%%%%%%%%%%%%%%%%%%%%%%%%%%%%%%%%%%%%%%%%%%%%%%%%%%%%%%%%%%%%%%%%
%%
%%   use this format to include an .eps figure into your paper
%%
\begin{figure}[htb]
\centering
\includegraphics[height=1.8in]{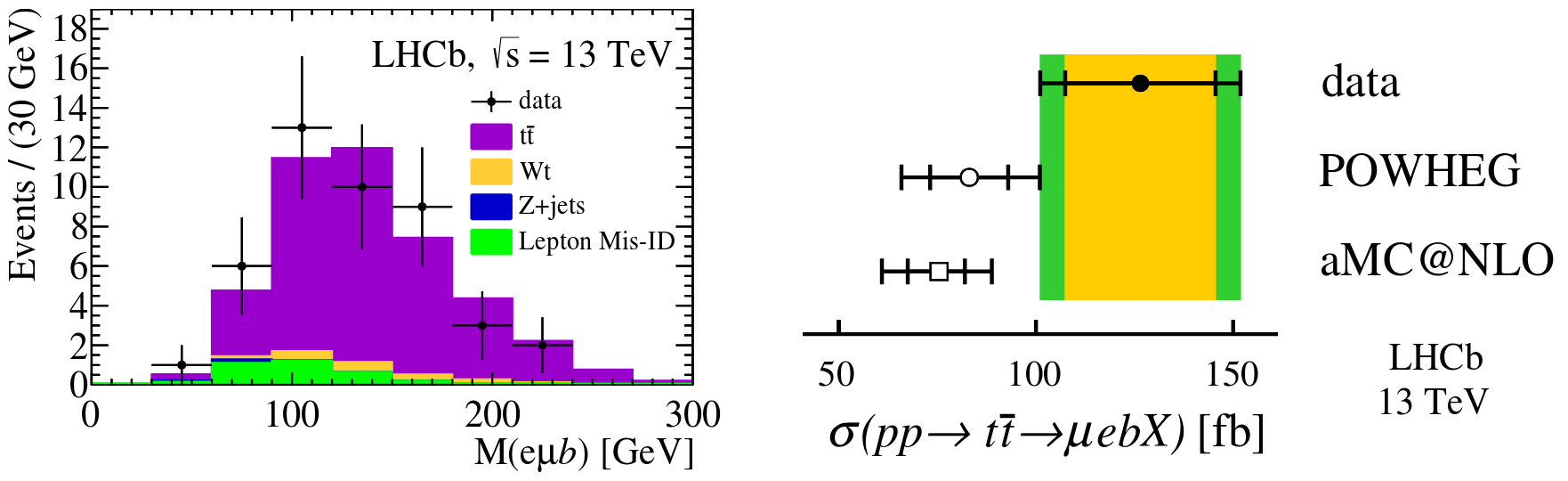}
\caption{Fit to invariant mass distribution of $\mu e b$ final state (left) with comparison of measured cross-section, where inner band represents statistical uncertainty, with theoretical predictions, where the inner band represents scale uncertainty (right) \cite{dilepton}.}
\end{figure}
%%%%%%%%%%%%%%%%%%%%%%%%%%%%%%%%%%%%%%%%%%%%%%%%%%%%%%%%%%%%%%%%%%%%%%%%%%%

The resulting top pair production cross-section in the fiducial region defined by: p$_{T}(l)>20\,$GeV; $2.0<\eta(\mu,e)<(4.5,4.25)\,$; $20<$p$_{T}(j)<100\,$GeV; $2.2<\eta(j)<4.2$; $\Delta$R$(l,j)>0.5$; $\Delta$R$(\mu,e)>0.1$, IP$_{l} < 0.04\,$mm and p$_{T}(j_{l})>5\,$GeV is:\\

\begingroup
\setlength{\tabcolsep}{.08\linewidth}
\begin{tabular}{r}
    $\sigma$($t\bar{t}$)[13\,TeV] \hspace{.2cm} = \hspace{.2cm} 126 $\pm$ 19 (stat) $\pm$ 16 (syst) $\pm$ 5 (lumi) fb \cite{dilepton}.
\end{tabular}
\endgroup

\vspace{.3cm}
\noindent These results are consistent with SM predictions and are in good agreement across jet and lepton kinematics. This highly pure statistically limited channel is set to benefit from the full Run$\,$II data set and future high statistics runs. With the increased data sets available in Run$\,$II and in future upgrades, subsequent measurements of top production at LHCb are expected to be alleviated of statistical limitations.

\end{document}